# Enhancing Management of Large-Scale Optical Networks through RFID Technology Integration

Xiaoying Zheng, Xingqi Xuan, Shilie Zheng, Xiaonan Hui*, Xianmin Zhang
*College of Information Science & Electronic Engineering*
*Zhejiang University*
Hangzhou, China
x.hui@zju.edu.cn

*Abstract*—Managing large-scale optical distribution networks is a daunting task. This paper introduces a novel solution using radio frequency identification (RFID) technology to transform the procedure we monitor and manage the complex optical network dumb resources (ONDR). By implementing and deploying removable RFID tag pairing based on the serial peripheral interface (SPI) communication protocol, the system identifies 30 pairs of tags within one second, even at densities of up to 5.1 tagged components per square inch of patch panel surface area. The integration of light-emitting diode (LED) navigation aids in indicating correctly matched interfaces, effectively addressing the complexities associated with large-scale fiber matching.

## I. INTRODUCTION

In recent years, large-scale optical communication systems have found widespread application in both industry and academia, where hundreds of thousands of optical connections intricately weave a complex cabling network. However, the sheer magnitude of optical connections presents a formidable challenge in managing optical network dumb resources (ONDR) effectively. It becomes evident that relying solely on manual processes to manage ONDR with absolute precision and minimal maintenance costs would be imprudent, given the scale and intricacy of such systems.

Researchers have explored different methods for optical fiber and connector identification, including quick response codes [1], and augmented reality [2]. But most of the endeavors encounter challenges related to operating complexity, environmental dependence, and information accessibility. As an alternative solution, radio frequency identification (RFID) technology is regarded as a key automated identification system [3]. RFID systems rely on low-power wireless backscatter communication [4], [5], [6], enabling low cost, compact size, and maintenance-free ONDR management.

In this paper, a UHF-RFID ONDR management system with inter-tag serial port communication to exchange tag ID and optic fiber connection logic is proposed. Information exchange between tags is accomplished through serial peripheral interface (SPI), effectively halving the number of identifications, and enhance positioning visibility with LED navigation.

## II. SYSTEM PRINCIPAL & DESIGN

RFID technology is the core of the ONDR management system, uniquely encoding and interconnecting each individual optical fiber and connector. Dumb resources are primarily distributed among passive transmission devices, such as splitters and switches. Fig. 1(a) illustrates a schematic scene depicting the management of optical fibers and connectors within a splitter.

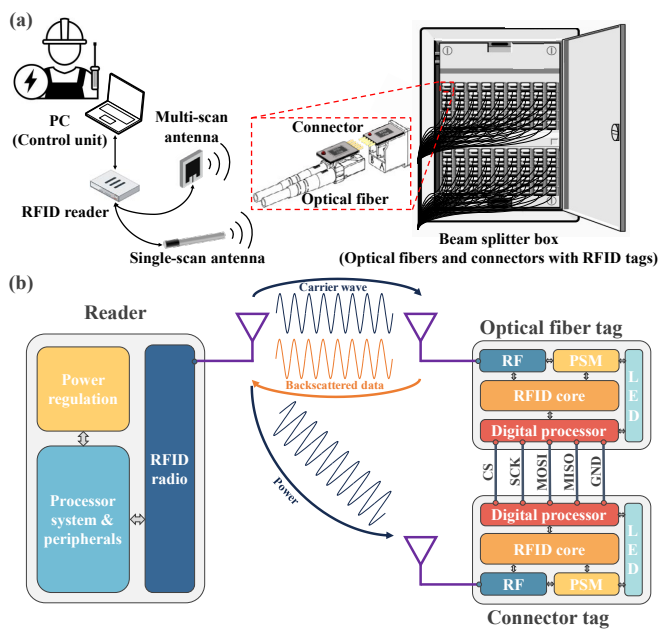

Fig. 1. Overview of the RFID ONDR management system. (a) The management scene of optical fibers and connectors in the splitter. (b) Block schematics of the RFID system showing wireless-powered communication network (WPCN).

Fig. 1(b) illustrates a passive RFID system is implemented for short-range wireless communication, comprising a reader, antenna, and passive tags. The passive tags can be conveniently deployed on ONDR and enable the identification of a wide range of objects through their unique electronic product code (EPC). For fiber connections, a single-scan antenna retrieves the tag EPC from the fiber, obtaining the target connector's EPC from the database. Subsequently, users scan the connector tags to identify the target connector, which activates its LED for navigation using a multi-scan antenna. To verify the accuracy of the optical fiber connections, users configure the fiber tags as the SPI master-mode and the connector tags as the SPI slave-mode. The SPI slave-module returns its EPC, enabling comparison with correct database records to ascertain connection accuracy, effectively halving the number of identifications.

## III. EXPERIMENTAL RESULTS

The implementation of a detachable 3D structural component facilitates swift installation of optical fibers and connectors, optimizing tag spatial distribution (Fig. 2).

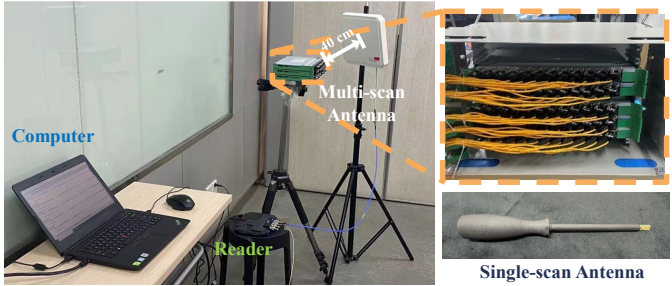

Fig. 2. Experimental setup of the RFID integrated fiber connectors with EPC exchange functionality. The orange dashed boxes show the RFID tags.

To enhance spatial efficiency and minimize interference between RFID tags, a self-designed helical antenna with a central frequency of 915 MHz is developed. The antenna features a height of 10 mm, a 5 mm diameter, and a 0.95 mm copper. Fig. 3(a) illustrates both simulated and measured reflection coefficients of the antenna, reaching −12.16 dB at 915 MHz with a bandwidth (< −10 dB) of 12.9 MHz (908.6−921.5 MHz). In practical application at 912 MHz, the reflection coefficient decreases to −13.74 dB, achieving a bandwidth of 15.5 MHz (906.0−921.5 MHz). Fig. 3(b) shows the antenna's radiation pattern.

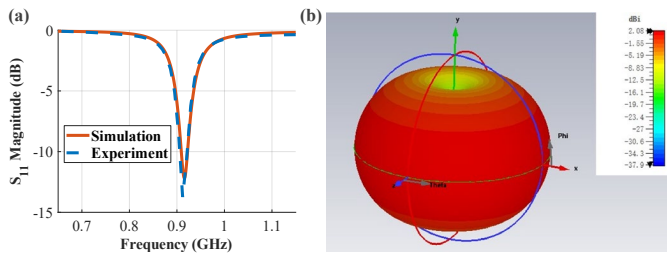

Fig. 3. Antenna performance. (a) Simulated and measured $S_{11}$ of the antenna. (b) Simulated radiation pattern of the antenna.

The read/write date rate can reach 119 samples per second (SPS), as shown in Fig. 4(a). In a standard indoor environment (25 °C), the reader's output power range is configured within 20−30 dBm, operating at a frequency range of 902−928 MHz. In this configuration, the tag's reception range spans from 0 to 125 cm with a maximum received signal strength indicator tolerance of −60 dBm. Fig. 2 presents 60 pairs of tags with a density of 5.1 tags per square inch. This setup was validated through 100 experiments, achieving a recognition efficiency of over 98%. Fig. 4(b) illustrates the distribution of time required for matching and verifying 30 pairs of tags, with 80% completing within 0.6 seconds.

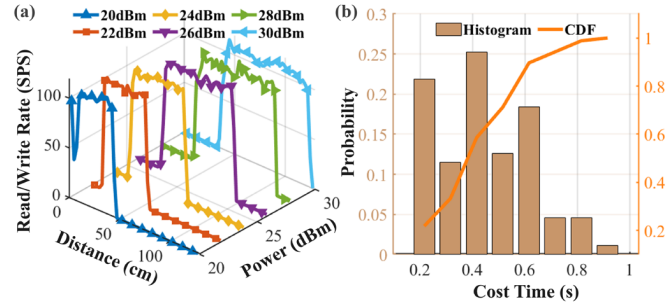

Fig. 4. Experimental output parameters of tags. (a) Read/Write rates of tags at different power levels and distances. (b) Distribution of time for 30 pairs of tags.

## IV. DISCUSSION AND CONCLUSION

This paper presents an innovating approach to revolutionize the monitoring and management of optical fiber connections by integrating RFID technology. By harnessing the passive and unique RFID tags, the proposed method enables the streamlined identification and management of ONDR. The quick-detachable structure and antenna design enhances spatial distribution for easy installation, making it a practical solution for efficient management of ONDR.

## ACKNOWLEDGMENT

The project is supported by the National Natural Science Fund for Excellent Young Scientists Fund Program (Overseas), the Ministry of Science and Technology of the People's Republic of China (2022YFB2903800), the Key R&D Program of Zhejiang (2023C03160) and the National Key Research and Development Program of China (2018YFB2201700). The authors gratefully acknowledge the support of Zhejiang University Education Foundation Qizhen Scholar Foundation.